\documentclass{PoS}

  \usepackage{amssymb}
  \usepackage{amsmath}
    \usepackage{amsfonts}
     \usepackage{epsfig}
      \usepackage{bm}

\usepackage[section]{placeins}

\usepackage{multirow}
\usepackage{ctable}
\usepackage{booktabs}
\usepackage{array}
\usepackage{tabularx}
\usepackage{xcolor}
\usepackage{pstricks}

\newcommand{\beq}{\begin{equation}}
\newcommand{\eeq}{\end{equation}}
\newcommand{\bea}{\begin{eqnarray}}
\newcommand{\eea}{\end{eqnarray}}
\newcommand{\beqa}{\begin{eqnarray}}
\newcommand{\eeqa}{\end{eqnarray}}

\newcommand{\bp}{\mbox{\boldmath $p$}}
\newcommand{\bq}{\mbox{\boldmath $q$}}

\newcommand{\be}{\mbox{\boldmath $e$}}

\def\lsim{\mathrel{\rlap{\lower4pt\hbox{\hskip1pt$\sim$}}
    \raise1pt\hbox{$<$}}}         
\def\gsim{\mathrel{\rlap{\lower4pt\hbox{\hskip1pt$\sim$}}
    \raise1pt\hbox{$>$}}}         

\definecolor{red}{rgb}{1,0,0}

\def\be{\begin{equation}}
\def\ee{\end{equation}}
\numberwithin{equation}{section}

\title{From deep-inelastic structure functions \\
to two-photon dilepton production \\ in proton-proton collisions}

\ShortTitle{From deep-inelastic structure functions}

\author{\speaker{Antoni Szczurek}\thanks{The work has been supported by the Polish National Science Center grant
DEC-2014/15/B/ST2/02528.}\\
Institute of Nuclear
Physics, Polish Academy of Sciences, Radzikowskiego 152,\\ PL-31-342 Krak{\'o}w, Poland\\
        E-mail: \email{antoni.szczurek@ifj.edu.pl}}

\author{Marta {\L}uszczak\\
Department of Theoretical Physics, University of Rzesz{\'o}w, PL-35-959 Rzesz{\'o}w, Poland\\
        E-mail: \email{luszczak@ur.edu.pl}}

\author{Wolfgang Sch\"afer\\
Institute of Nuclear
Physics, Polish Academy of Sciences, Radzikowskiego 152,\\ PL-31-342 Krak{\'o}w, Poland\\
        E-mail: \email{wolfgang.schafer@ifj.edu.pl}}

\abstract{
We compare two different approaches used for calculating
cross sections for the two-photon $p p \to l^+ l^- X$ process. 
In one of the approaches photon is treated as a collinear parton in 
the proton. 
In the second approach a recently proposed $k_T$-factorization
method is used. 
In this presentation we discuss sensitivity of the results to 
the choice of structure function parametrization and experimental cuts 
in the $k_T$-factorization approach.
We compare results of our calculations with recent experimental data
for dilepton production and find that in most cases
the contribution of the photon-photon mechanism is rather small.
We discuss how to enhance the photon-photon contribution.
We also compare our results to those of recent measurements of 
exclusive and semi-exclusive $e^+ e^-$ pair production with certain 
experimental data by the CMS collaboration.
}

\FullConference{XXIV International Workshop on Deep-Inelastic Scattering and Related Subjects\\
		11-15 April, 2016\\
		DESY Hamburg, Germany}

\begin{document}

\section{Introduction}

 In this presentation we show our recent results published
in \cite{Luszczak} where we have considered 
$p p \to p p l^+ l^-$ (exclusive) and 
$p p \to l^+ l^-$ (semiexclusive, with proton dissociation)
double photon fusion processes in the proposed somewhat earlier
$k_t$-factorization approach 
\cite{daSilveira:2014jla}.
In our presentation at DIS2016 we have focussed on
the relation of
the cross section for the charged lepton pair production 
with the dependence of the deep-inelastic structure functions
$F_2$, that are the input for our approach, on $x$ and $Q^2$.

\begin{figure}
\begin{center}
\includegraphics[width=3.5cm]{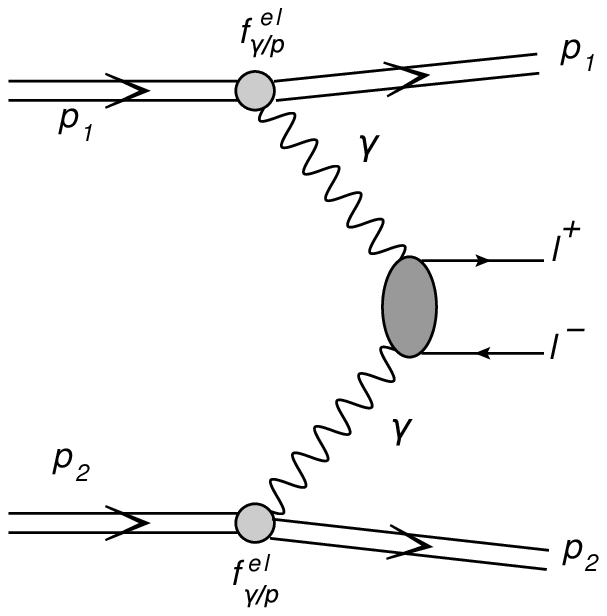}
\includegraphics[width=3.5cm]{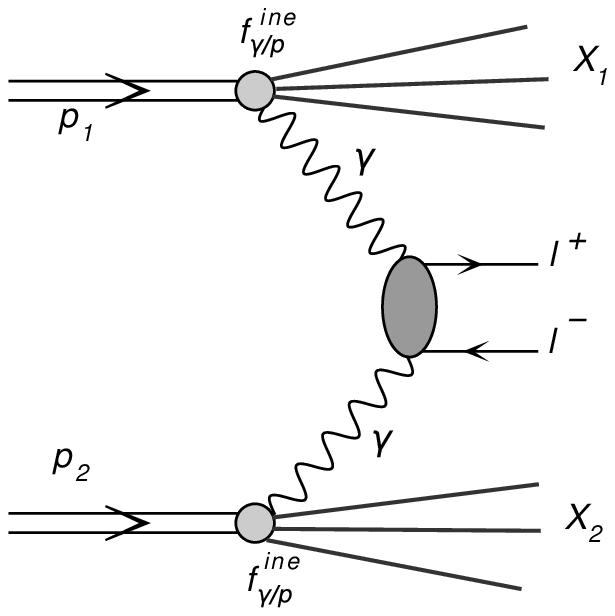}
\includegraphics[width=3.5cm]{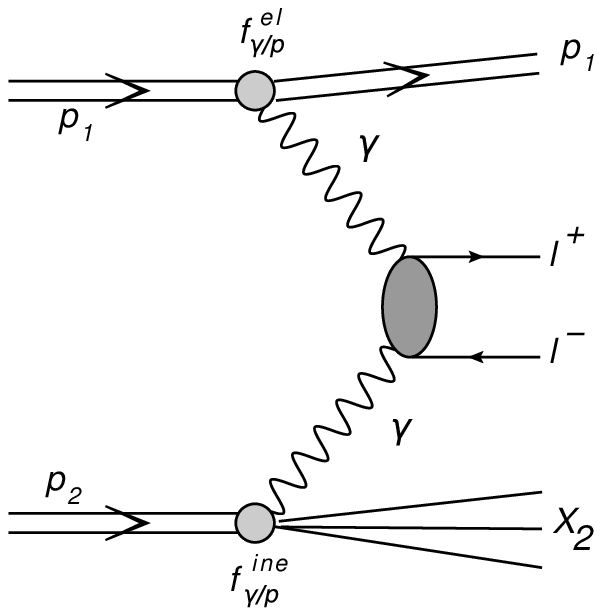}
\includegraphics[width=3.5cm]{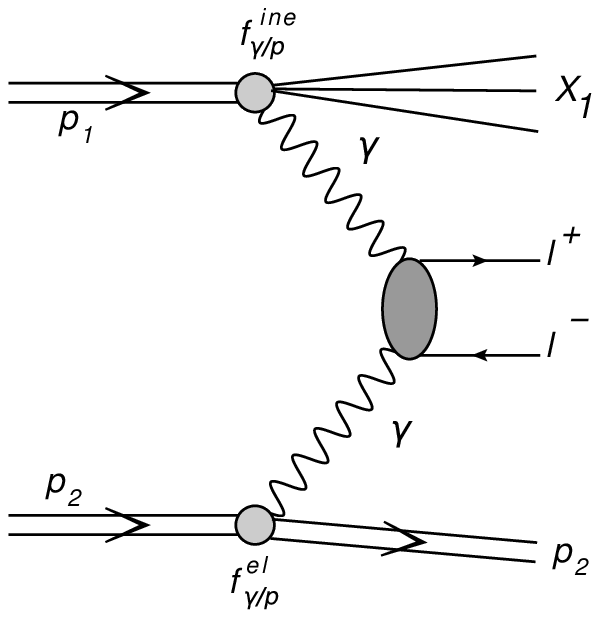}
\end{center}
{\caption \small
Different mechanisms of two-photon production of dileptons
included in \cite{Luszczak}.
}
\label{fig:mechanisms}
\end{figure}

The main mechanisms of the dilepton production considered
in \cite{Luszczak} are shown in Fig.\ref{fig:mechanisms}.
The considered mechanism has the same final state as the dominant
Drell-Yan mechanism. For $k_t$-factorization approach to the Drell-Yan
see \cite{Szczurek:2008ga,NNS2013,Baranov:2014ewa,SS2016}.
In our recent paper \cite{Luszczak} we discussed in detail the
photon-photon mechanism.

\section{Basic formulae}

In collinear approximation the cross sections are calculated as:
\begin{equation}
\frac{d \sigma^{\gamma_{in} \gamma_{in}}}{d y_1 d y_2 d^2p_t} = \frac{1}{16 \pi^2 {\hat s}^2}
x_1 \gamma_{i}(x_1,\mu^2) \; x_2 \gamma_{j}(x_2,\mu^2) 
\overline{|{\cal M}_{\gamma \gamma \to l^+l^-}|^2} \; ,
\end{equation}
where i,j=el,in and $f_i$ are photon PDFs.
The elastic photon fluxes are calculated using 
the Drees-Zeppenfeld parametrization, 
where a simple parametrization of 
nucleon electromagnetic form factors was used.



In the $k_t$-factorization approach the differential cross section can be
written as:
{\small
\begin{eqnarray}
 {d \sigma^{(i,j)} \over dy_1 dy_2 d^2\bp_1 d^2\bp_2} &&=  
\int  {d^2 \bq_1 \over \pi \bq_1^2} {d^2 \bq_2 \over \pi \bq_2^2}  
{\cal{F}}^{(i)}_{\gamma^*/A}(x_1,\bq_1) \, 
{\cal{F}}^{(j)}_{\gamma^*/B}(x_2,\bq_2) 
{d \sigma^*(p_1,p_2;\bq_1,\bq_2) \over dy_1 dy_2 d^2\bp_1 d^2\bp_2} 
\, , \nonumber \\ 
\label{eq:kt-fact}
\end{eqnarray}
}
where i,j=el,ine and ${\cal F}_k$ are unintegrated fluxes of photons.
As shown in \cite{Luszczak} the unintegrated fluxes can be expressed in
terms of the (deep-inelastic) structure functions $F_2(x,Q^2)$.

%
%

\section{Numerical results}

In our studies in \cite{Luszczak} we have used a few different 
parametrizations of the proton structure function $F_2$ taken 
from the literature:
\begin{itemize} 
 \item ALLM \cite{Abramowicz:1991xz,Abramowicz:1997ms}. This
   parametrization gives a very good fit to $F_2$ in most of 
   the measured region.
\item FJLLM \cite{Fiore:2002re}. This parametrization explicitly
  includes the nucleon resonances and gives an excellent fit of the CLAS data.
 \item BDH \cite{Block:2014kza}. This parametrization concentrates on
   the low-$x$, or high mass region. It features a Froissart-like
   behaviour at very small $x$. 
  \item SY \cite{Suri:1971yx}. This paramerization of Suri and Yennie
    from the early 1970's does not include QCD-DGLAP evolution. It is
    still today often used as one of the defaults in the LPAIR event generator.
 \item SU \cite{Szczurek:1999rd}. A parametrization which concentrates
   to give a good description at rather small and intermediate $Q^2$ at 
   not too small $x$.
\end{itemize}

We also show $F_2$ calculated from the CTEQ6L parametrization 
\cite{Pumplin:2002vw}.

\begin{figure}[!h]
\begin{center}
   \includegraphics[width = 0.4\textwidth]{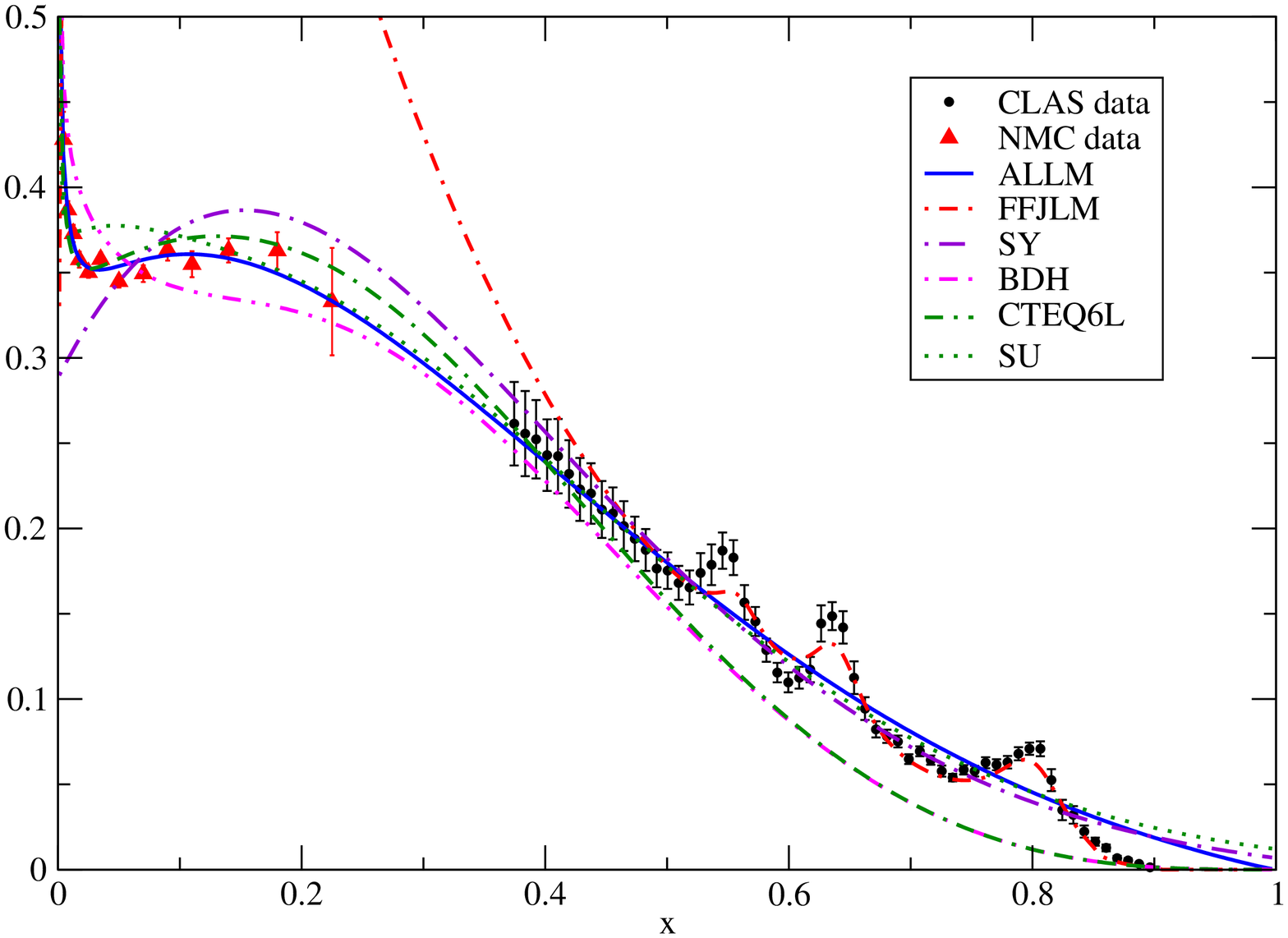}
   \includegraphics[width = 0.4\textwidth]{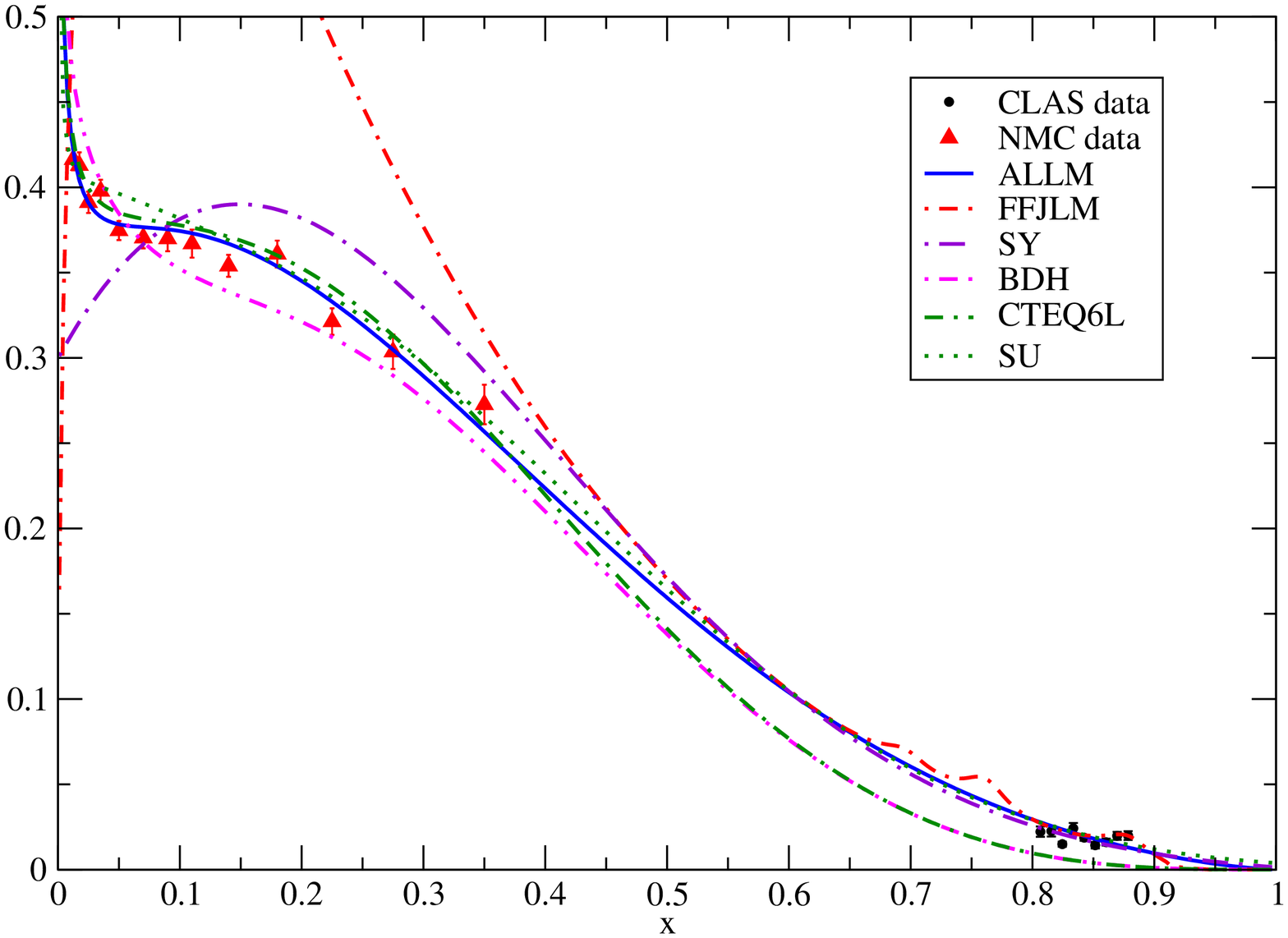}
\end{center}
   \caption{The proton structure function $F_2(x,Q^2)$ as a function 
of $x$ for 
$Q^2 = 2.5 \, \rm{GeV}^2$(left), and 
$Q^2= 4.5 \, \rm{GeV}^2$(right).
Shown are results for different parmetrizations available in the literature.
 }
  \label{fig:F2}
 \end{figure}

In Fig.\ref{fig:F2} we show only two examples of the proton structure 
function $F_2(x,Q^2)$ obtained from the various parametrizations 
at $Q^2 = 2.5, 4.5  \, \rm{GeV}^2$ 
as a function of Bjorken-$x$. 


It is surprising that the old Suri-Yennie \cite{Suri:1971yx} fit,
still gives a reasonable description of $F_2$ except of very small $x$.
For explicit account of resonances we recommend to use the
Fiore et al. \cite{Fiore:2002re}, but care has to be taken
to stay within the resonance region, as the quality of the fit beyond
this region quickly deteriorates. 
The overall best description appears to be given by the ALLM 
\cite{Abramowicz:1991xz,Abramowicz:1997ms} fit.

In Ref.\cite{Luszczak} we have compared our calculations with
measured dilepton data 
\cite{PHENIX,LHCb-CONF_2012,ATLAS_2014_low-mass,ATLAS_2013_high-mass,ISR}.
Here we show only a few examples.

\begin{figure}
\begin{minipage}{0.45\textwidth}
 \centerline{\includegraphics[width=1.0\textwidth]{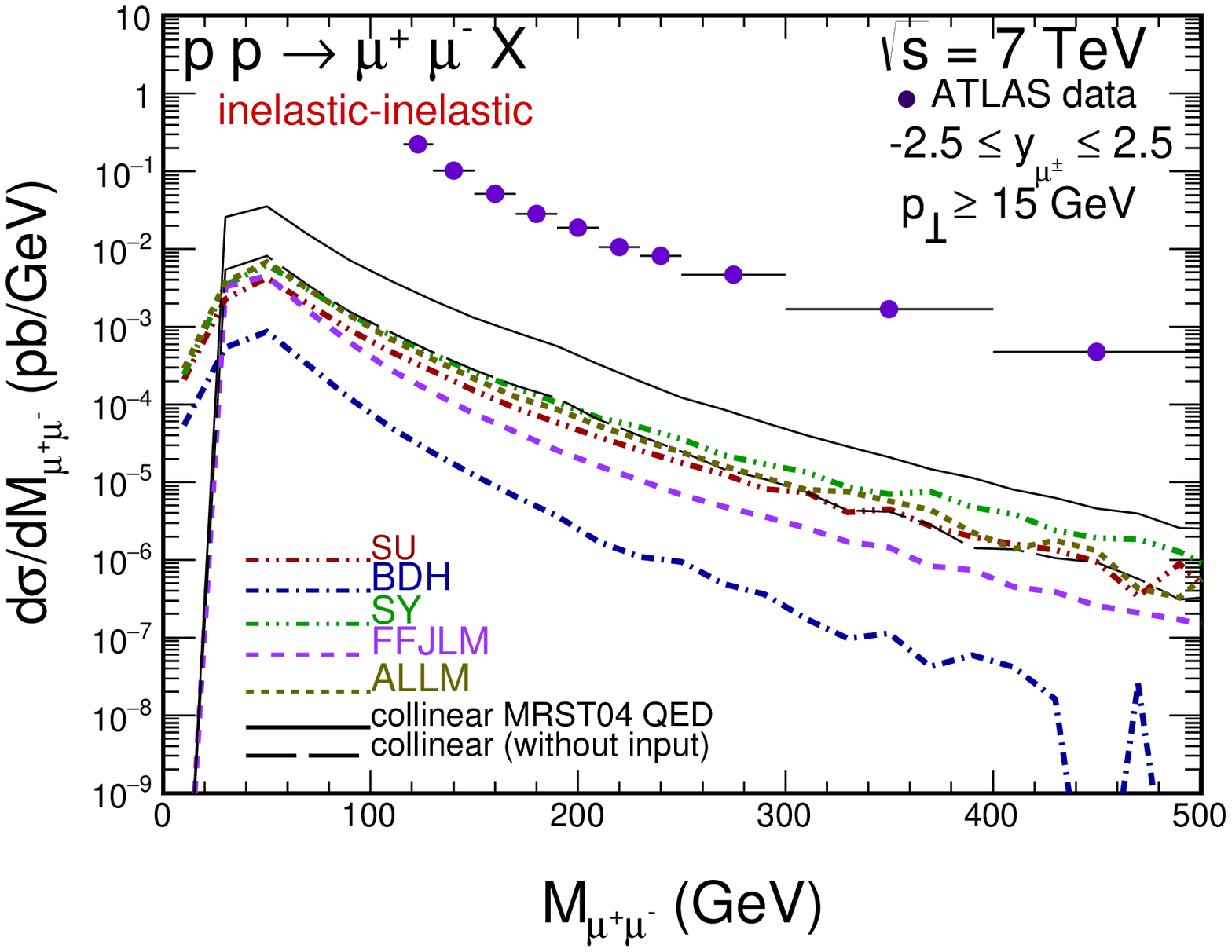}}
\end{minipage}
\hspace{0.2cm}
\begin{minipage}{0.45\textwidth}
 \centerline{\includegraphics[width=1.0\textwidth]{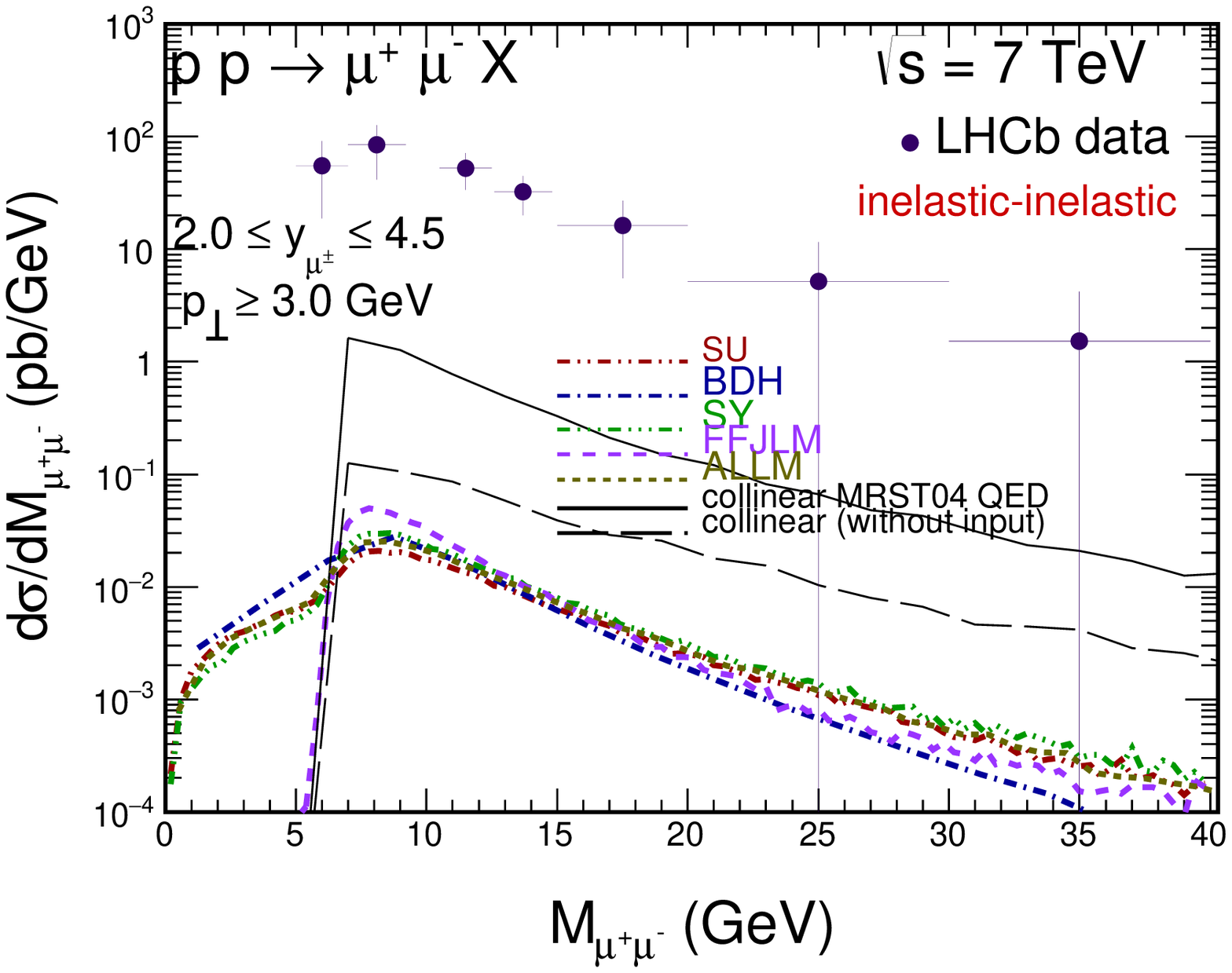}}
\end{minipage}
   \caption{The inelastic-inelastic contribution to dilepton invariant mass 
distributions for 
ATLAS (left) and LHCb (right) experiments 
for different structure functions.}
 \label{fig:dsig_dMll_ine_ine}
\end{figure}

Most of the experiments for the dilepton production concentrate on
determination of dilepton invariant mass distributions. 
In Fig.\ref{fig:dsig_dMll_ine_ine} we show invariant mass distributions
of dilepton pairs produced in the photon-photon inelastic-inelastic 
mechanism for kinematical conditions relevant for different experiments.
We show results obtained with the different parametrizations of the structure
functions known from the literature. Surprisingly the different
structure functions give quite different results.
For completeness in some cases we also show 
the result obtained in the collinear approach with the MRST2004(QED) 
photon distribution \cite{Martin:2004dh} with (solid black line)
and similar one when ignoring the initial input (long-dashed black line). 
The result obtained within the collinear approach with the 
MRST2004(QED) distribution is much above the results obtained within 
the $k_t$-factorization approach.
In our opinion this is mainly related to the large input photon distribution
at the initial scale $Q_0^2$ = 2 GeV$^2$. 
If the input is discarded (long-dashed black line) the collinear
result is similar to the results obtained within the $k_T$-factorization.
The inelastic-inelastic contribution gives only a small fraction
of the measured cross section for most experimental conditions 
(ATLAS,LHCb). 

\begin{figure}
\begin{minipage}{0.45\textwidth}
 \centerline{\includegraphics[width=1.0\textwidth]{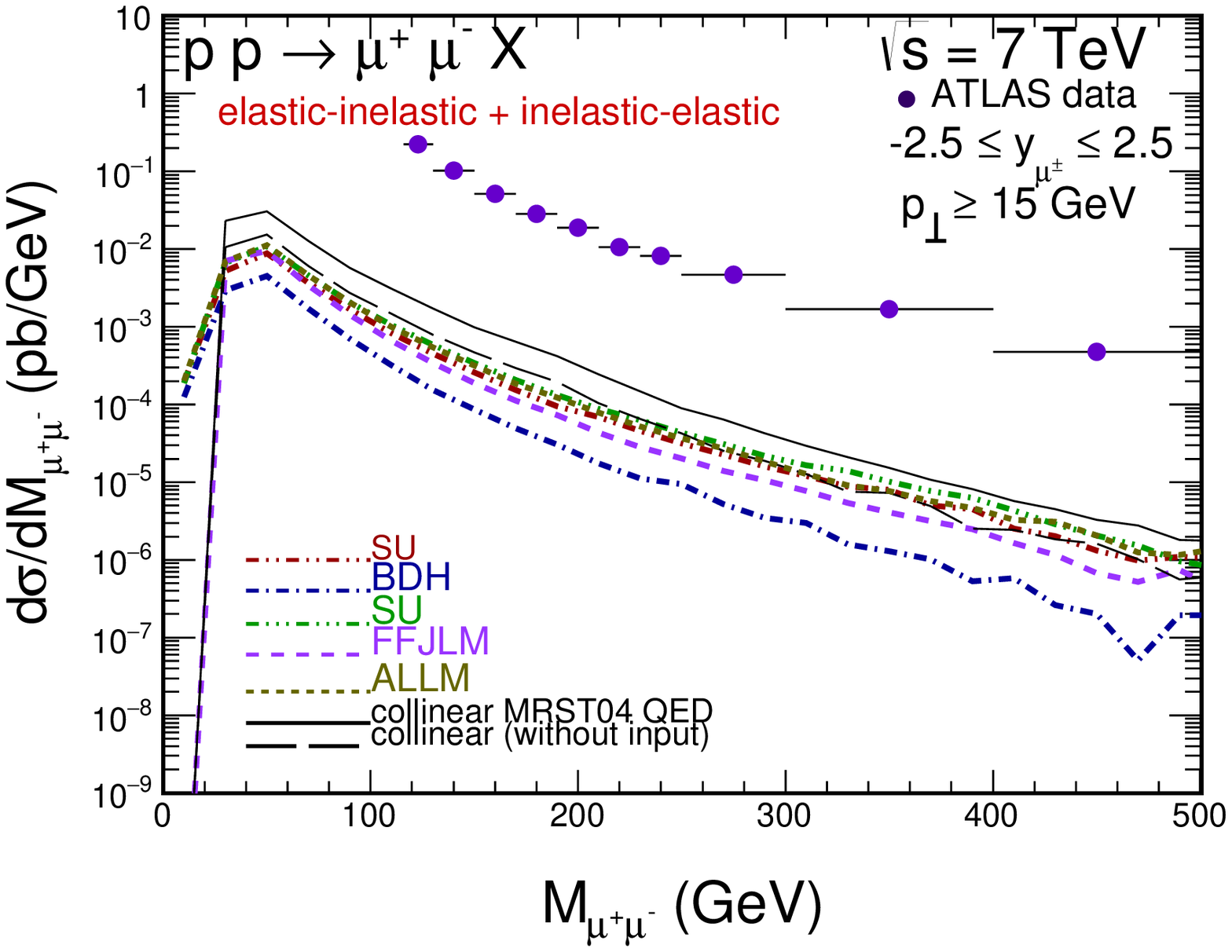}}
\end{minipage}
\hspace{0.2cm}
\begin{minipage}{0.45\textwidth}
 \centerline{\includegraphics[width=1.0\textwidth]{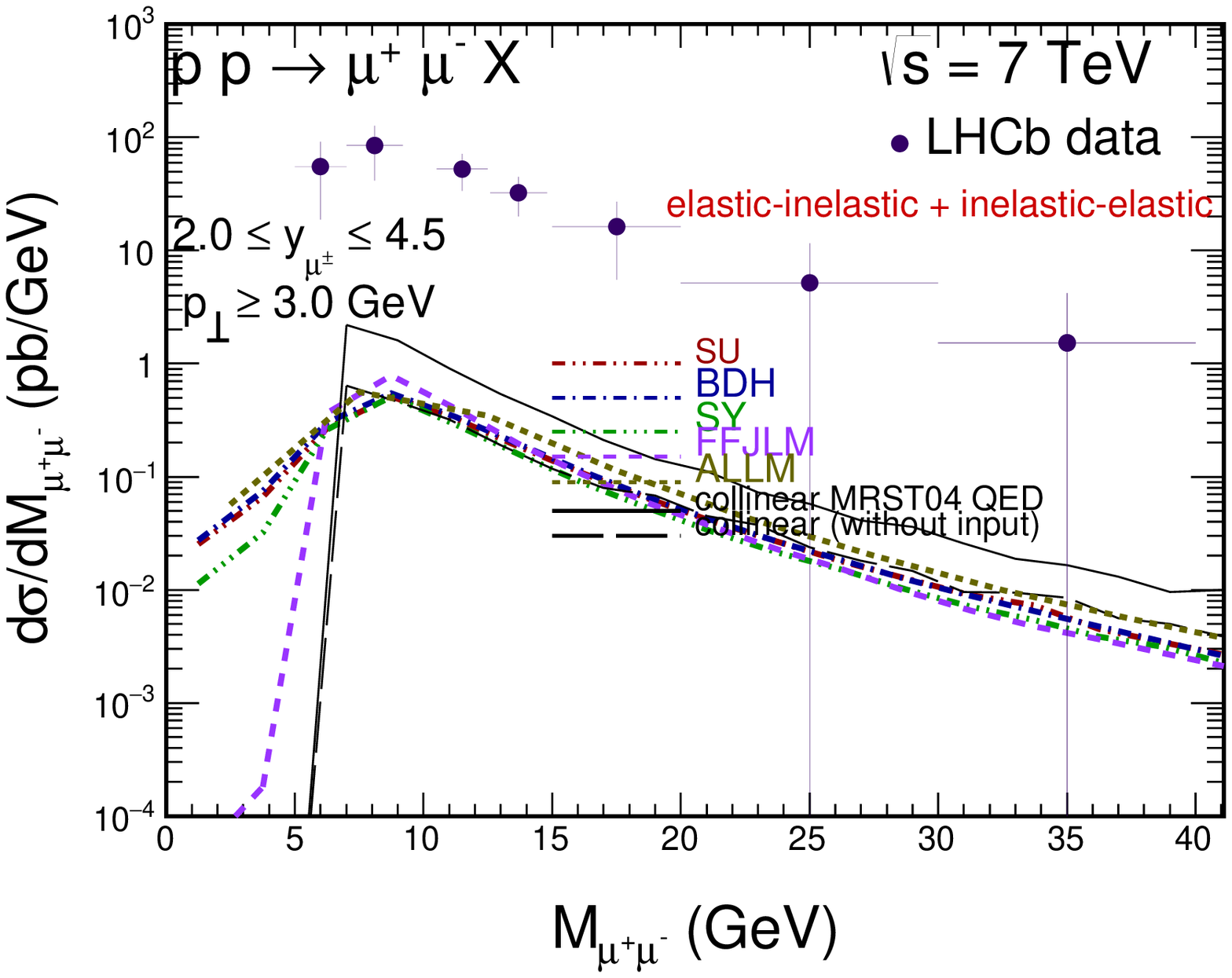}}
\end{minipage}
   \caption{The (elastic-inelastic)+(inelastic-elastic) contribution 
to dilepton invariant mass distributions for 
ATLAS (left) and LHCb (right) experiments 
for different structure functions.}
 \label{fig:dsig_dMll_ela_ine}
\end{figure}

In Fig.\ref{fig:dsig_dMll_ela_ine} we show dilepton invariant mass 
distributions for elastic-inelastic and inelastic-elastic (added
together) contributions.
As for inelastic-inelastic contribution the results strongly depend
on the parametrization of the structure functions used.
The spread of results for different $F_2$ from the literature
is now somewhat smaller than in the case of 
inelastic-inelastic contributions where the structure functions enter
twice.
As for the double inelastic case we also show a result for
the collinear approach.
The mixed components give similar contribution to the dilepton
invariant mass distributions as the inelastic-inelastic one.

\begin{figure}
\begin{center}
\includegraphics[width=5cm]{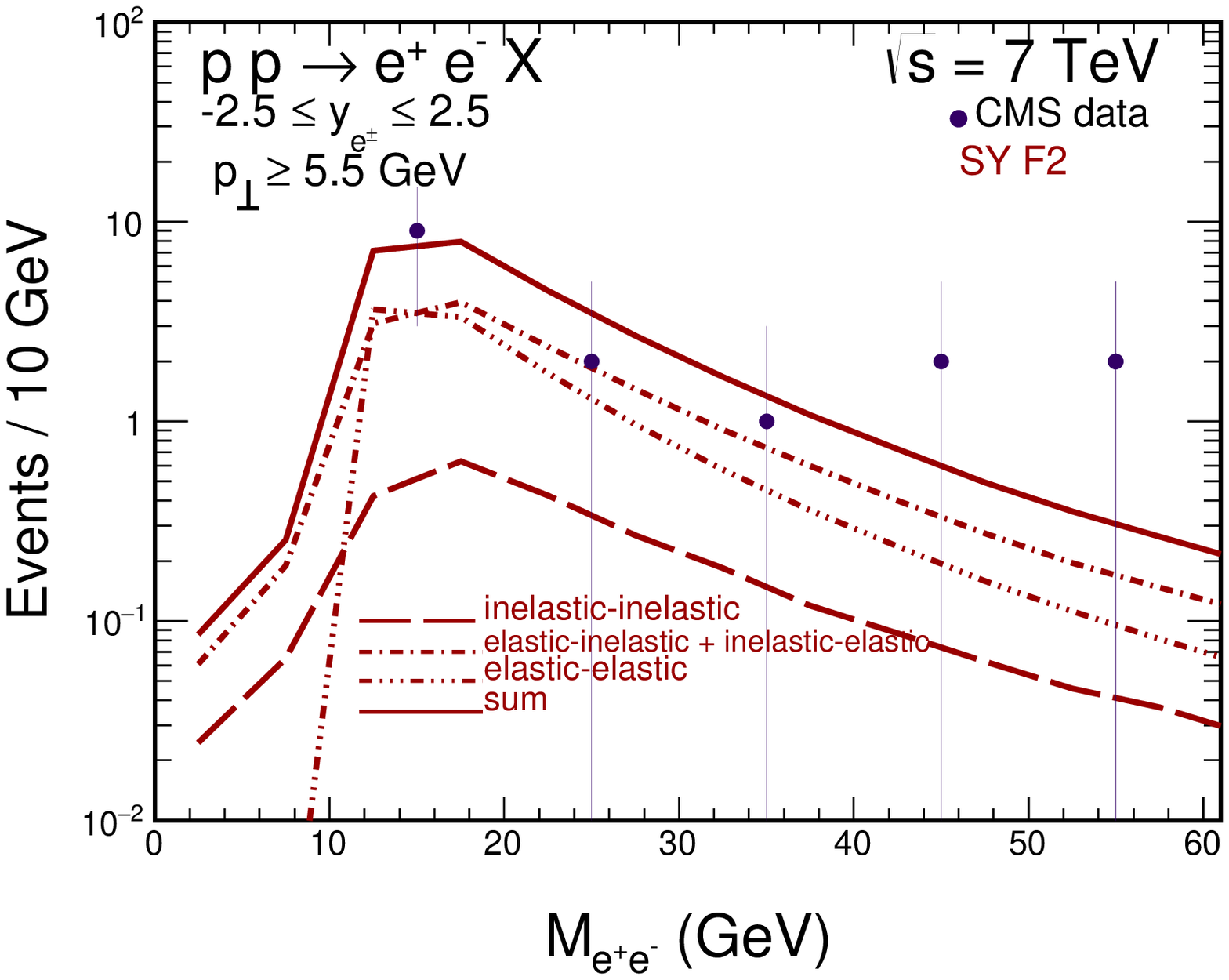}
\includegraphics[width=5cm]{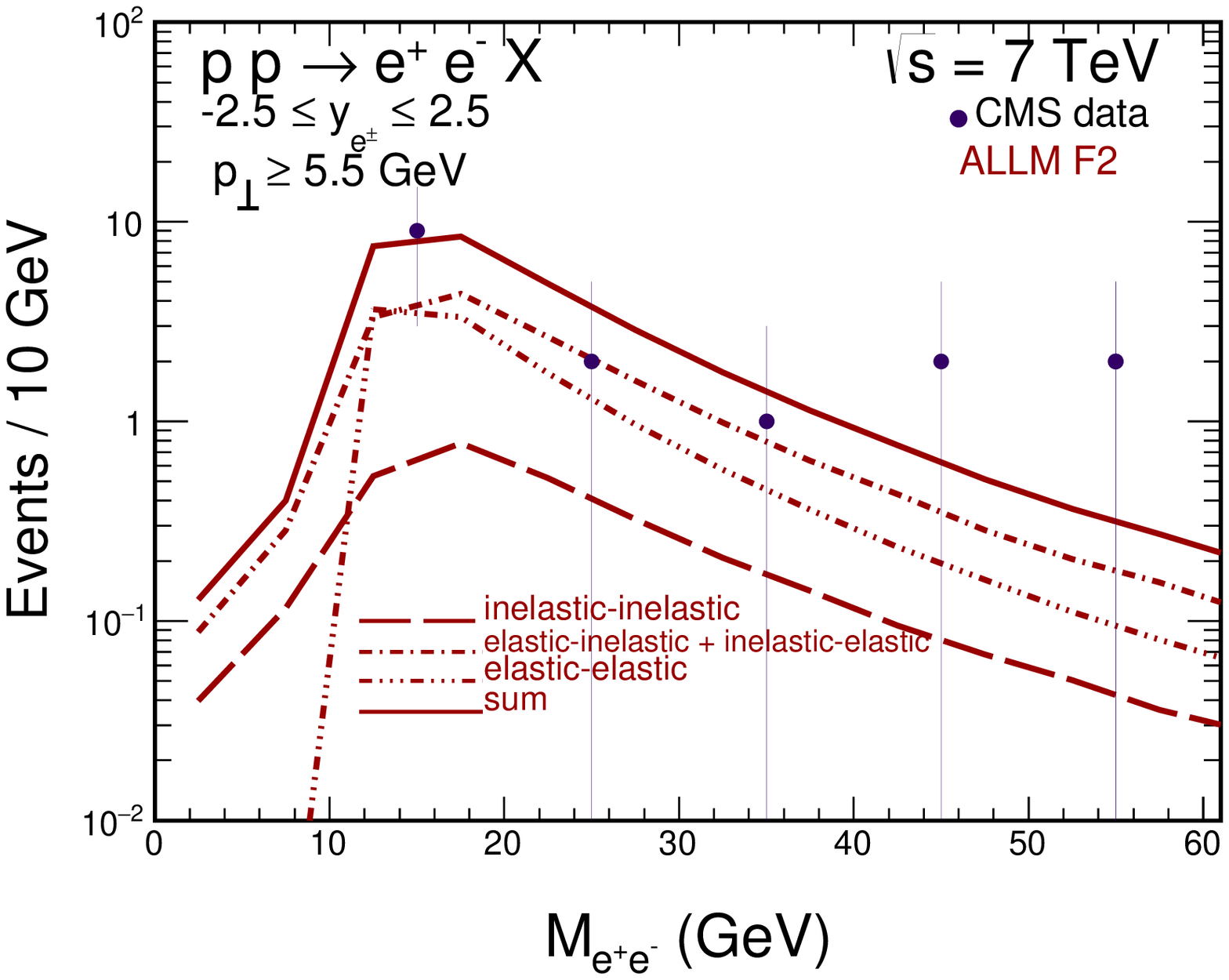}
\end{center}
\caption{
Number of events per invariant mass interval for the CMS experimental
cuts for SY (left) and ALLM (right) structure functions.
The experimental data points are from Ref.\cite{Chatrchyan:2012tv}.
}
\label{fig:CMS_dN_dMll}
\end{figure}

\begin{figure}
\begin{center}
\includegraphics[width=5cm]{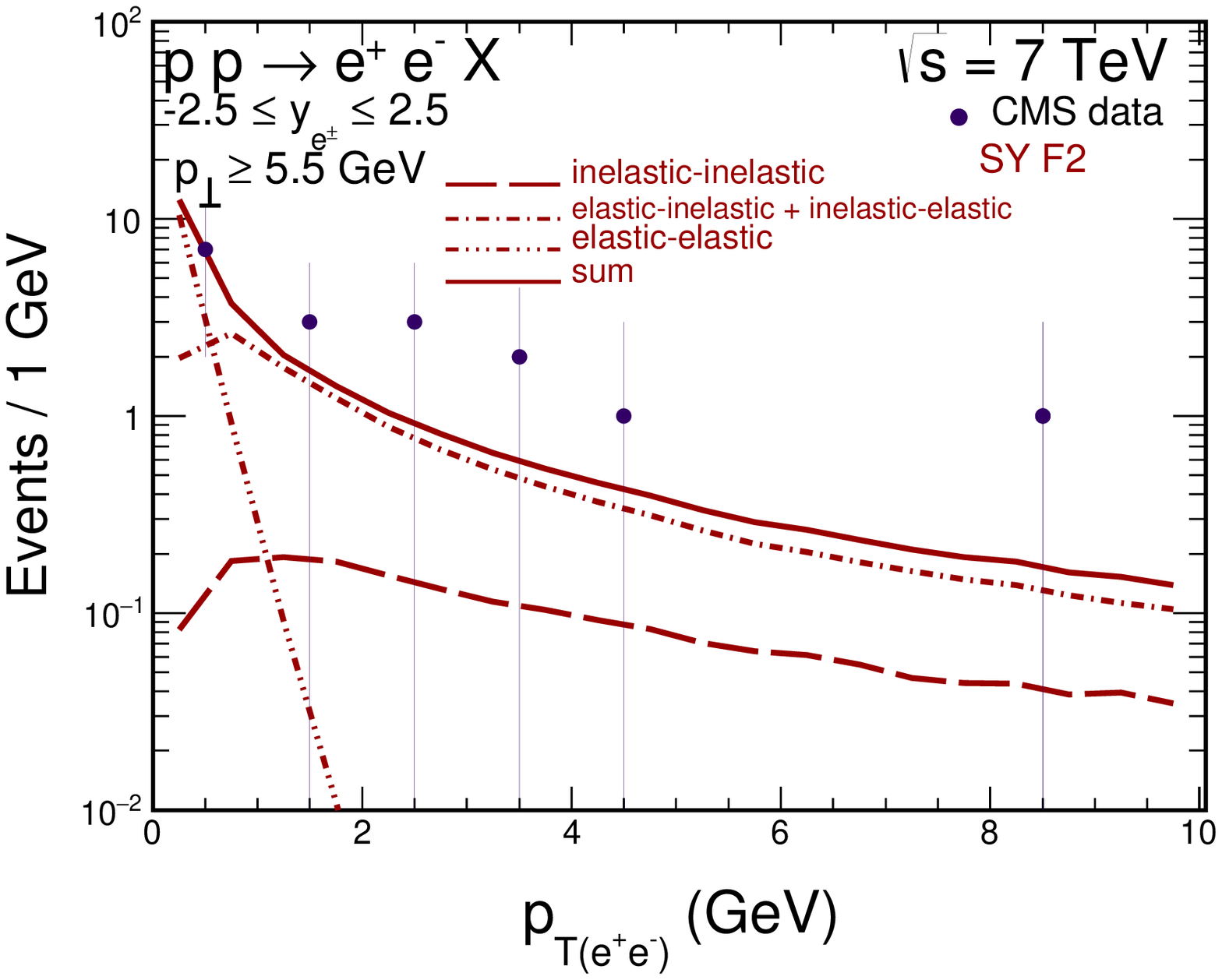}
\includegraphics[width=5cm]{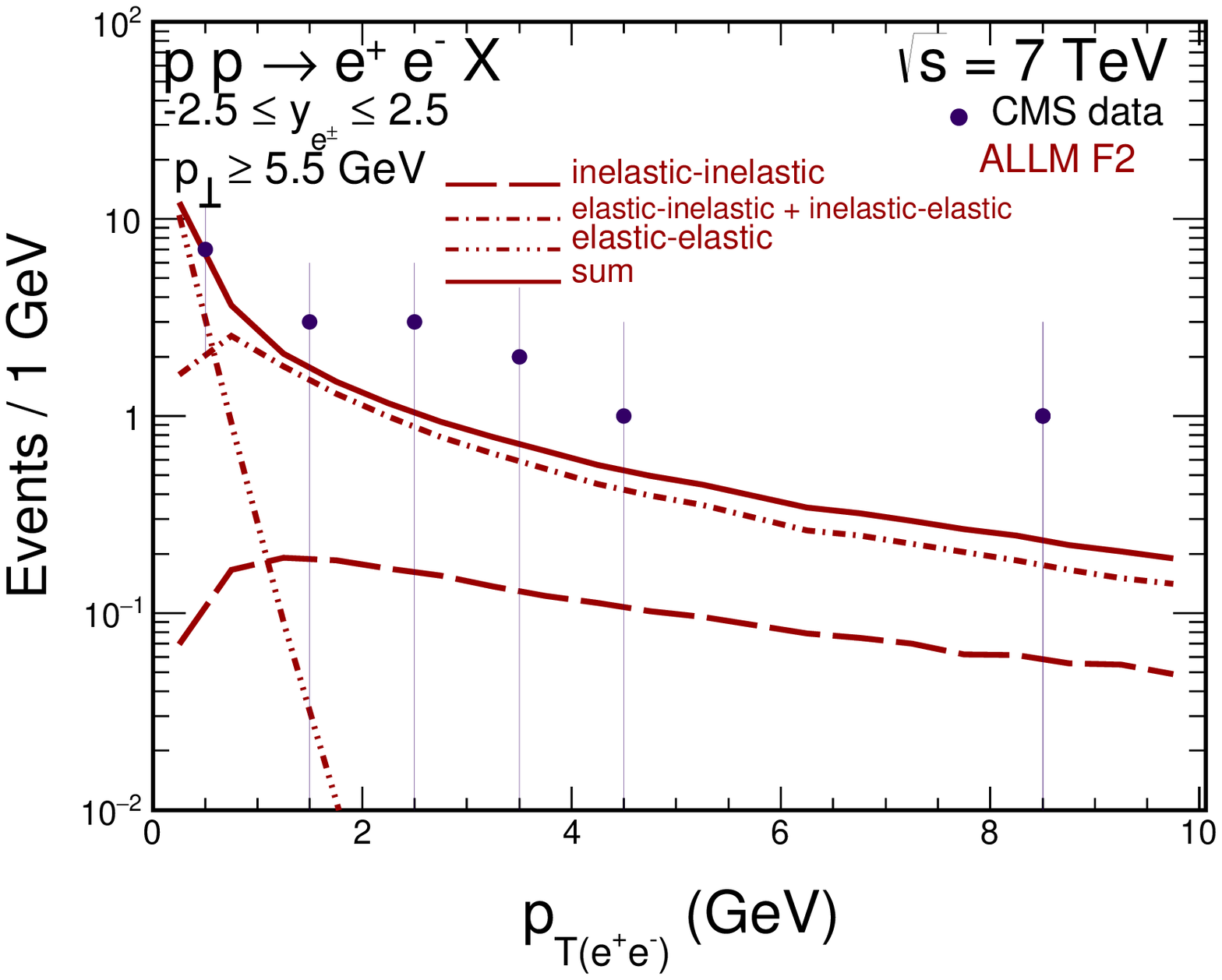}
\end{center}
\caption{
Number of events per pair transverse momentum interval for the CMS
experimental cuts for SY (left) and ALLM (right) structure functions.
The experimental data points are from Ref.\cite{Chatrchyan:2012tv}.
}
\label{fig:CMS_dN_dptsum}
\end{figure}

\begin{figure}
\begin{center}
\includegraphics[width=5cm]{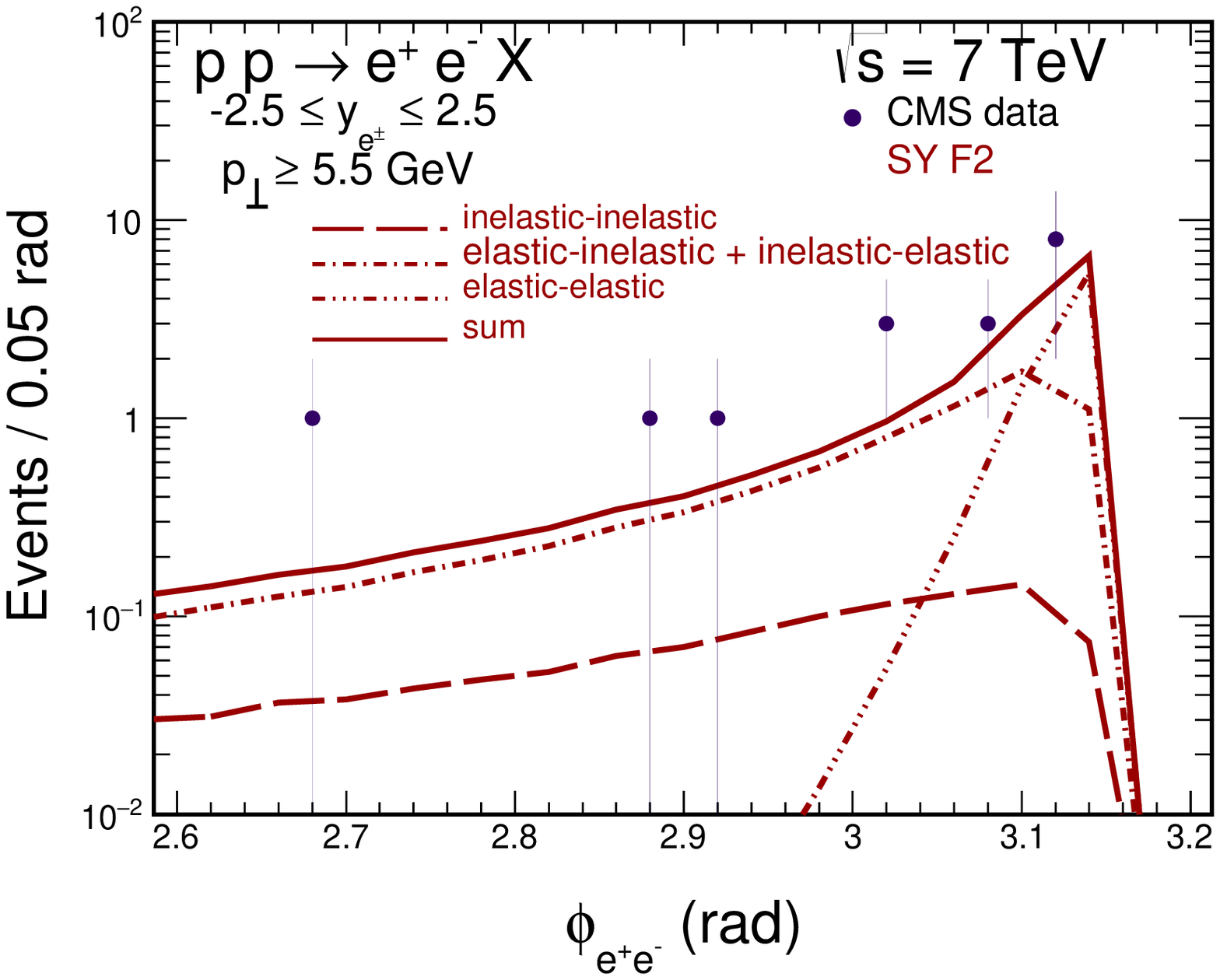}
\includegraphics[width=5cm]{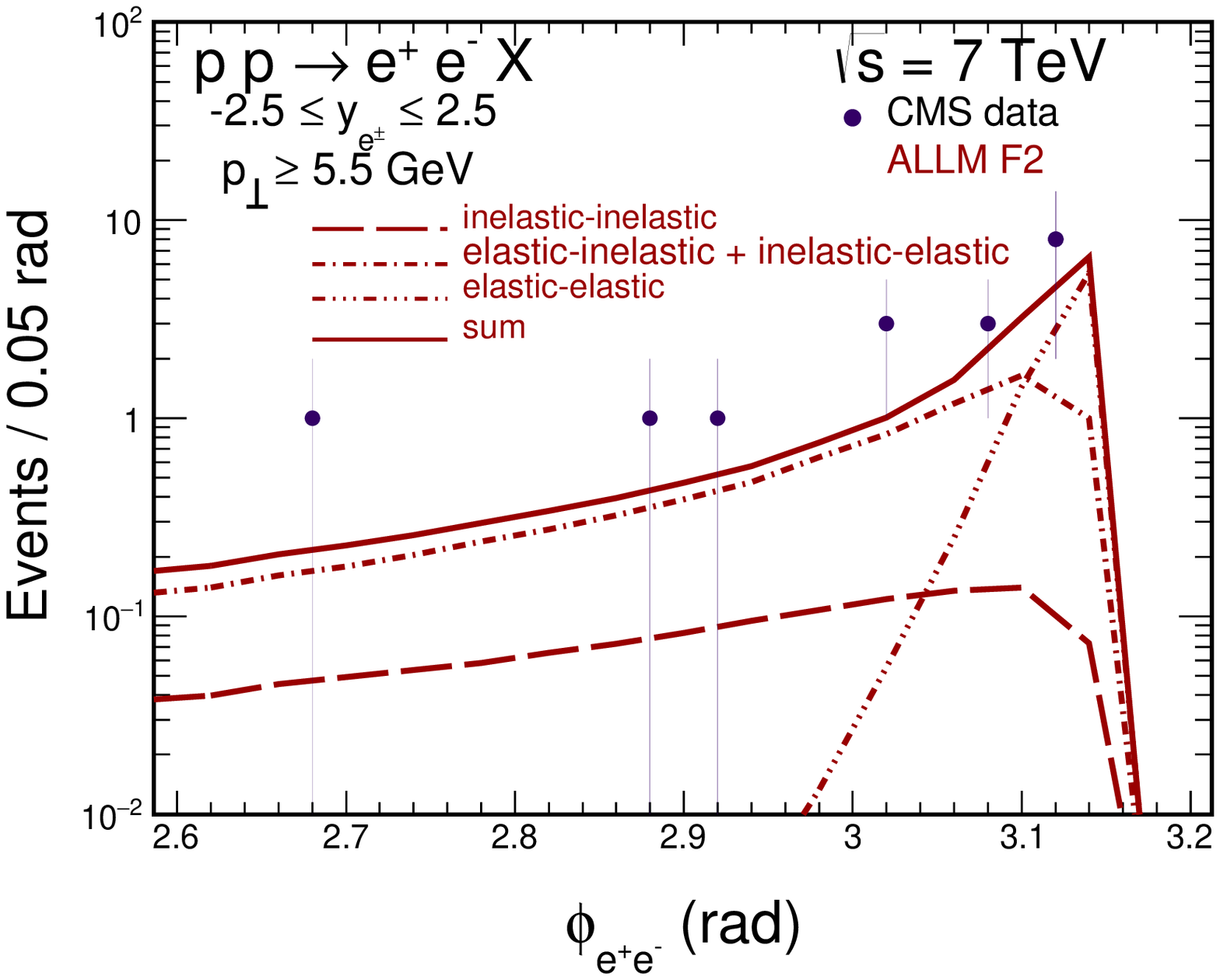}
\end{center}
\caption{
Number of events per pair relative azimuthal angle interval
for the CMS experimental cut for SY (left) and ALLM (right) structure
functions. The experimental data points are from Ref.\cite{Chatrchyan:2012tv}.
}
\label{fig:CMS_dN_dphi}
\end{figure}

In most of the cases considered so far Drell-Yan processes dominate 
\cite{NNS2013,Baranov:2014ewa,SS2016}.
The two-photon processes are interesting by themselves.
Can they be measured? 
In order to reduce the Drell-Yan contribution and relatively enhance 
the two-photon contribution one can impose an extra condition on lepton
isolation. First trials have been done by the CMS collaboration
\cite{Chatrchyan:2012tv}. In their analysis an extra lepton isolation
cuts were imposed in order to eliminate the dominating
Drell-Yan component.
In Figs.
\ref{fig:CMS_dN_dMll},\ref{fig:CMS_dN_dptsum},\ref{fig:CMS_dN_dphi}
we show our results for two different (SY and ALLM) parametrizations 
of the structure functions for distributions in dimuon invariant mass,
in transverse momentum of the pair and in relative azimuthal angle
between $\mu^+ \mu^-$. 
SY and ALLM parametrizations give almost 
the same contributions to all the distributions considered.
In the first evaluation we have taken into account integrated luminosity
of the experiment ($L$ = 63.2 pb$^{-1}$) as well as experimental
acceptances given in Ref.\cite{Chatrchyan:2012tv}.
Rather good agreement with the low statistics CMS experimental data is
achieved without including any extra corrections due to absorption
effects.
It may mean that the absorption effects are small or alternatively 
that a contamination of the Drell-Yan contribution is still 
not completely removed.
Both effects should be therefore studied in detail in a future.

\section{Conclusions}

We summarize our studies in \cite{Luszczak} as follows:

\begin{itemize}

\item Two different approaches (collinear and $k_t$-factorization)
      for $\gamma \gamma \to l^+ l^-$ processes 
      were discussed and compared.

\item Strong dependence on the structure function input
      in the $k_t$-factorization approach were found.

\item Semi-exclusive contributions with proton dissociation is large
      (this may be interesting lesson for other processes
       such as e.g. the $p p \to p pJ/\psi$ reaction).

\item Photon-photon contribution is rather small compared 
      to Drell-Yan contribution but is important in
      precision calculations. 

\item Reasonable description of the CMS data with 
      isolated electrons was achieved
      (recently also ATLAS obtained similar result).

\item The regions of the arguments of the structure function $F_2$ 
      important for the discussed $\gamma \gamma \to l^+ l^-$ process   
      was identified.

\item So far only collinear approach was applied to  
      $p p \to (\gamma \gamma) \to W^+ W^- X Y$ processes
      which is important in searches for Beyond Standard Model effects.

\end{itemize}


\end{document}